\documentclass[preprint,preprintnumbers,amsmath,amssymb,aps]{revtex4}

\usepackage[dvips]{graphicx}
\usepackage{amssymb,amsfonts,amsmath}
\usepackage{multirow}
\usepackage{ulem}

\begin{document}

\title{The Effect of Network Topology on the Stability of Discrete State Models of Genetic Control}

\author{Andrew Pomerance}
 \email{pomeranc@umd.edu}
\author{Edward Ott}
\author{Michelle Girvan}
\author{Wolfgang Losert}
\affiliation{%
Institute for Research in Electronics and Applied Physics \\
University of Maryland, College Park\\
College Park, MD, 20752
}%
\begin{abstract}
Boolean networks have been proposed as potentially useful models for genetic control.  An important aspect of these networks is the stability of their dynamics in response to small perturbations.  Previous approaches to stability have assumed uncorrelated random network structure.  Real gene networks typically have nontrivial topology significantly different from the random network paradigm.  In order to address such situations, we present a general method for determining the stability of large Boolean networks of any specified network topology and predicting their steady-state behavior in response to small perturbations. Additionally, we generalize to the case where individual genes have a distribution of `expression biases,' and we consider non-synchronous update, as well as extension of our method to non-Boolean models in which there are more than two possible gene states.  We find that stability is governed by the maximum eigenvalue of a modified adjacency matrix, and we test this result by comparison with numerical simulations.  We also discuss the possible application of our work to experimentally inferred gene networks. 
\end{abstract}

\maketitle

\section{\label{sec:introduction}Introduction}
Boolean networks have been extensively investigated as a model for genetic control of cells \cite{Kauffman1969, Kauffman1993}.  In this model, each gene is represented by a node of a network, and each node has one of two states: on -- i.e., producing (`expressing') its target protein -- or off.  Directed links between genes indicate that one gene influences the expression of another.  This can correspond to the expressed protein directly binding to DNA and modulating the transcription of a gene or to other signaling pathways that modulate DNA transcription.  In the standard Boolean network model, the system evolves in discrete timesteps ($t = 0, 1, 2,...$), and at each step the state of every node is simultaneously updated according to some function of its inputs.  This function approximates the action of activators (proteins which act to increase expression of a given gene) or inhibitors (proteins which act to reduce expression).  While this model might seem to be an oversimplification considering the complex kinetics involved in all steps of a transcription pathway, experimental evidence suggests that real biological systems are, in some cases, reasonably well-approximated by Boolean networks \cite{Albert2003}. 

In 1969, S.A. Kauffman \cite{Kauffman1969} introduced a type of Boolean network known as an $N-K$ network.  In this model, there are $N$ nodes each having exactly $K$ input links, and the nodes from which these input links originate are chosen randomly with uniform probability.  We refer to the number of input (output) links to (from) a node as the in-degree (out-degree) of that node.  At any given time $t$, the system state can be represented as an $N$-vector whose $i$th component $\sigma_i(t)$ is either zero or one, where $i = 1,2,...,N$.  There are $2^N$ possible states.  The function determining the time evolution at each node is defined by a random, time-independent, $2^K$-entry truth table.  Since this is a finite, deterministic system, there is always an attractor: eventually, the system must return to a previously visited state (finiteness), after which the subsequent dynamics will be the same as for the previous visit (determinism).  These attractors can be fixed points or periodic orbits.  Using the Hamming distance between two states (i.e., the number of nodes for which the $\sigma_i(t)$ disagree) as the distance measure, the system exhibits both what is termed a `chaotic' (or unstable) regime, where the distance between typical initially close states on average grows exponentially in time, as well as a stable regime, where the distance decreases exponentially.  Between the two there is a `critical' regime.  (Here by `close' we mean that the Hamming distance is small compared to $N$.)

As a model of genetic control, these attractors have been postulated to represent a specific pattern of protein expression which defines the cell's character \cite{Kauffman1969}.  In single-celled organisms, these attractors might be taken to correspond to different cell states (growing, dividing, starving, heat- or pH-shocked).  In multi-cellular organisms, different cell types (muscle, nerve, liver, etc.) have different expression patterns, and, within each type, a cell could be in a variety of states (resting, `activated,' dividing, etc.) that each correspond to different expression patterns.  Boolean network approximations have been successful in predicting the gene expression time sequence of the segment polarity gene network in \textit{Drosophilia}, a model for embryonic development where individual cells turn specific proteins on and off in patterns that guide the growth of certain organs and structures \cite{Albert2003}.  Since the protein expression pattern of the cell is modeled from the state of the corresponding Boolean network, the question of the stability of the network then becomes important: do small perturbations in the expression pattern, due perhaps to chemical fluctuations, die out quickly, returning the cell to its original state, or do they quickly grow, pushing the cell into another state?  The purpose of this paper is to examine the stability of network dynamics in the context of discrete state models of gene networks.

One motivation for the consideration of dynamical stability is its possible relevance to cancer.  Specifically, we hypothesize that dynamical instability of a gene network might be a causal mechanism contributing to the occurence of some cancers.  We emphasize that this hypothesis is distinct from the previous hypothesis of `genomic instability' as a cause of cancer \cite{GenomicInstab1}.  In particular, genomic instability has been defined\footnote{As defined in the glossary of \textit{Nature Genetics}}  as `the failure to transmit an accurate copy of the entire genome from one cell to its two daughter cells.'  In contrast, the instability we refer to is that of the \textit{dynamics} of a \textit{given} gene network, and we use the term `dynamical network instability' (DNI) to distinguish this condition.  We speculate that DNI might arise from mutations and that, once established, as cells divide, DNI could lead to widely varying gene expression patterns from cell to cell.  We emphasize that DNI implies that this variation would arise even in the absence of further mutation.  That is, similar to the concept of chaos in continuous-state dynamical systems (e.g., \cite{OttBook}), DNI causes exponential sensitivity of typical system trajectories to small changes, which we speculate may lead to many different outcomes in the course of cell division.  Recent microdissection results indicate wide variations in gene expression patterns even for nearby cells within the same cancerous tissue \cite{w1}. This variability provides a basis for understanding why cancer can adapt and evade treatment \cite{w2}.

Another motivation for our study is the argument, put forward by Kauffman \cite{Kauffman1993}, that evolution favors gene networks that are on the border between stability and instability \cite{Critical1, Critical2, Critical3, Critical4}. Whether or not our cancer hypothesis or Kauffman's stability-border hypothesis holds, the question of dynamical stability of such networks is crucial to their understanding and use as models.

While previous works have addressed the question of dynamical network stability in simple, specific types of random networks (e.g. $N-K$ nets), in this paper we address the question of dynamical network stability for general network topology and node attributes.  We also consider nonsynchronous update and extend the considerations to non-Boolean models allowing for the possibility of nodes having more than two states.  Thus our work provides a potentially enhanced framework for modeling and using the discrete state network paradigm.  In particular, we consider how our network stability considerations can be employed on experimentally derived gene networks.

In the original $N-K$ nets as proposed by Kauffman, the truth table output governing node dynamics was randomly chosen with on and off having equal probability.  Subsequently, it was shown that if the truth table output was biased such that $p$ denotes the probability of randomly assigning an off output, the transition between the stable and chaotic regimes depends on $p$ \cite{Derrida1986}.  We term $p$ the `expression bias.'  Additionally, networks with a distribution of in-degrees, but no in-/out-degree correlation, have been considered in \cite{Luque1995, Luque1997, Fox2001, Aldana2003}, and it has been shown that the nodal in-degree average, $\langle K^{in} \rangle$, suffices to determine the stability.  (Here $\langle \cdot \rangle$ indicates average of a nodal quantity over all nodes.)  Specifically, the critical average number of connections, $K_c$, governing this transition is 
\begin{equation}
K_c  = 1/[2p(1-p)],
\label{eq:rbn_phase}
\end{equation}
where the network is stable for $\langle K^{in} \rangle < K_c$, unstable for $\langle K^{in} \rangle > K_c$, and critical for $\langle K^{in} \rangle = K_c$.  Aldana and Cluzel \cite{Aldana2003} considered the consequences of Eq. \eqref{eq:rbn_phase} in the case of networks with scale-free topology \cite{Barabasi1999}, i.e., the probability distribution $P(K^{in})$ (or $P(K^{out})$) that a randomly chosen node has in-degree $K^{in}$ (out-degree $K^{out}$) is a power-law: $P(K) \propto K^{-\gamma}$.  (Since every out-link for a node is an in-link for some other node, $\langle K^{in}\rangle = \langle K^{out} \rangle$; thus the result is unchanged whether it is the in- or out-degree that has power-law scaling.)  

Recently, some authors have noted, but not numerically tested, a generalization of Eq. \eqref{eq:rbn_phase} that takes into account nodal correlations between the in-degree and out-degree characterized by the joint $K^{in}-K^{out}$ degree distribution function $\tilde{P}(K^{in}, K^{out})$.  In this case, the critical transition occurs at \cite{Lee2008}
\begin{equation}
\label{eq:corr_phase}
\frac{\langle K^{in}K^{out} \rangle}{\langle K \rangle} = \frac{1}{2p(1-p)}.
\end{equation}

We emphasize that Eq. \eqref{eq:rbn_phase} was derived in the annealed approximation (see later discussion) for networks with a given in- or out-degree distribution $P(K)$ and with the complimentary links completely random, and that Eq.  \eqref{eq:corr_phase} uses only the additional information contained in the nodal in-/out-degree correlation.  Furthermore, all nodes (`genes') were taken to have the same $p$ value.  However, gene networks, in common with real networks occurring across a broad range of applications, can be expected to deviate substantially from the above simple network model.  Examples of network properties that could make previous analyses of network stability inapplicable are assortativity \cite{Newman2002} (the tendency for highly connected nodes to prefer or avoid linking to other highly connected nodes) and community structure \cite{Girvan2004} (the existence of highly connected, sparsely interconnected subgraphs), two properties that are not captured in the degree distributions.  Additionally, these properties may have biological implications.  For example, a recent paper \cite{Wang2007} examined gene interaction networks from cancerous tissue and found significant community structure, as well as positive correlation between the in-degree and out-degree of nodes; additionally, protein interaction networks have been shown to exhibit significant disassortativity \cite{Newman2002, ProteinNet}.  Furthermore, for modeling purposes, it might be important to allow the expression bias $p$ to vary from node to node (as an extreme example, so-called housekeeping genes \cite{Housekeeping} have a predominant tendency to be on, corresponding to low $p$, unlike other genes).  In this paper, we derive and test the stability criterion for large networks with arbitrary network topology and heterogeneous expression biases.  In particular, our theory evaluates the stability of any given network with its specific topology (i.e., its adjacency matrix $A$ defined subsequently), and by its node-specific expression biases.  We show that stability is determined by the largest eigenvalue of a modified adjacency matrix, and we numerically test this criterion.

With respect to real gene networks, the synchronous update at integer times ($t = 0, 1, 2, ...$) used in the above models represents an additional deviation from the real situation, where chemical kinetics and transport processes can be expected to introduce non-trivial dynamics.  As a partial step toward remedying this (and to make Boolean approximations suitable for atmospheric and geophysical processes), Ghil and Mullhaupt \cite{Ghil} consider a generalization in which $t$ is a continuous variable and $\sigma_i(t)$ depends on $\sigma_j(t-\tau_{ij})$, where $\tau_{ij}$ is a delay time that can be different for each link from $j$ to $i$.  The original formulation (e.g., in Refs. \cite{Kauffman1969, Derrida1986, Luque1995, Luque1997, Fox2001, Aldana2003}) corresponds to $\tau_{ij} = 1$ for all $i, j$.  We will argue and numerically confirm that the criterion determining the stability/instability border of this generalization of the Boolean network model is the same as that for the synchronous update models.  

In addition to nonsynchronous update, another generalization of Boolean networks that we will examine is models in which each node $i$ is allowed to have one of $S_i$  possible discrete states (e.g., for $S_i = 3$, we label the states $\sigma_i \in \{0,1,2\}$, and for Boolean networks $S_i = 2$ for all $i$).  This model may be closer to the behavior of actual cells, and models with multiple states can be related to certain piece-wise ODE models of transcription \cite{ODE, Multivalue2}.  The general model using arbitrary, multivalued truth tables has been previously treated in the special case of $N-K$ networks with all nodes having the same number of possible states $S$.  In the case where each possible state is equally likely, the critical number of inputs is \footnote{Unpublished results by Sole, RV, Luque, B, and Kauffman, S.}
\begin{equation}
\label{eq:multiphase}
K_c = \frac{1}{1-1/S}.
\end{equation}

The applicability of our work to any specific network and set of node-wise expression biases may be of particular interest in situations where experimental data provide the possibility of estimating a gene network and expression biases.  Such information could be used as input to our method which could give an indication of the stability of a given experimentally-derived network.  The possibility that such analyses may be feasible becomes more and more likely with the rapid technological advances in obtaining new types of high-quality, quantitative data useful for deducing gene networks.  For example, such analyses could be used to address the hypothesis that dynamical instability of gene networks is connected with the occurence of cancer.

\section{\label{sec:theory}Model}

Deterministic Boolean networks are formally defined by a state vector $\Sigma(t) = [\sigma_1(t) \sigma_2(t) ... \sigma_N(t)]^T$, where $\sigma_i \in \{0, 1\}$, and a set of update functions $f_i$, such that
\begin{equation}
\label{eq:definition}
\sigma_i(t) = f_i(\sigma_{k(i,1)}(t-1), \sigma_{k(i,2)}(t-1), ... ),
\end{equation}
where $k(i,1), k(i,2),..., k(i, K^{in}_i)$ denote the indices of the $K^{in}_i$ nodes that input to node $i$; we denote this set of nodes by $\mathcal{K}_i = \{k(i,j)| j=1,2,..,K^{in}_i\}$.  The update function $f_i$ is defined at each node $i$ by specifying a truth table whose outputs are randomly populated.  Previous analytic results assumed a constant expression bias for all nodes; however, we allow that, in the truth table for node $i$, output entries are randomly assigned zero with probability $p_i$ or one with probability $1-p_i$.  In the case of uniform expression bias, we drop the subscript and use the notation $p \equiv p_i$.

We consider the interaction structure of this system as a graph where the nodes represent individual elements of the state vector, and a directed edge is drawn from node $j$ to node $i$ if $j \in \mathcal{K}_i$.  An adjacency matrix $A$ is defined in the usual way: a matrix entry $A_{ij}$ is one if there is a directed edge from node $j$ to node $i$ and zero otherwise.

The stability of a large Boolean network is defined by considering the trajectories resulting from two close initial states, $\Sigma(t)$ and $\tilde{\Sigma}(t)$. To quantify their divergence, the Hamming distance of coding theory is used: $h(t) = \sum_{i=1}^N |\sigma_i(t) - \tilde{\sigma_i}(t)|$.  If the network is stable, on average $h(t) \rightarrow 0$ as $t \rightarrow \infty$.  In unstable networks, $h(t)$ quickly increases to $O(N)$, while a `critical' network is at the border separating stability and chaos.

In order to study the stability of an $N-K$ Boolean network, Derrida and Pomeau \cite{Derrida1986} considered an annealed situation and calculated the probability that, after $t$ steps, a node state is the same on two trajectories that originated from initially close conditions.  (This calculation was later generalized to variable in-degree \cite{Luque1995, Luque1997, Fox2001}, and joint degree distribution in \cite{Lee2008}.) In Derrida and Pomeau's `annealed' situation, at \textit{each} time step $t$ the truth table outputs and the network of connections are randomly chosen.  The actual situation of interest, however, is the case of `frozen-in' networks, where the truth table and network of connections are fixed in time.  It has been commonly assumed that analytical results obtained for the annealed case are a good approximation to the frozen-in case (e.g., Refs. \cite{Derrida1986, Luque1995, Luque1997, Fox2001}).  We also adopt this view in a modified form, and we will test its predictions with numerical simulations.

The randomization of the network of connections at each time step while keeping the degree distribution fixed carries the implicit assumption that there is no additional dynamically relevant structure in the frozen network other than that contained in the joint degree-distribution $\tilde{P}(K^{in}, K^{out})$. To avoid this assumption, we obtain theoretical results for a different annealing protocol, which we term `semi-annealed.'  In this semi-annealed procedure, we keep the network fixed (i.e., the adjacency matrix $A$ does not change with time), and we envision randomly assigning the output entries of the truth table of each node $i$ at every time $t$ according to the time-independent expression bias $p_i$ assigned to node $i$.  We then imagine tracking the probability that individual node states $\sigma_i(t)$ and $\tilde{\sigma}_i(t)$ differ over time with an $N$-dimensional difference vector, whose components are $y_i(t) = \langle\langle |\sigma_i(t) - \tilde{\sigma_i}(t)| \rangle\rangle $, where $\langle\langle \cdot \rangle\rangle$ denotes an average over every possible small initial perturbation.  Here by `every possible small initial perturbation' we mean all perturbations for which a small fraction $\epsilon$ of the states are flipped.  Additionally, we define the `sensitivity' $q_i$ as the probability that the output of $f_i$ changes when given two different input strings, similar to the `average sensitivity' of Ref. \cite{sensitivity}.  In the case of completely random Boolean functions,
\begin{equation}
\label{eq:sensitivity}
q_i = 1 - \bigg(p_i^{2}+(1-p_i)^{2}\bigg) = 2p_i(1-p_i).
\end{equation}
Thus, similar to Ref. \cite{Derrida1986}, we can write the update equation for $y_i$ as
\begin{equation}
y_i(t) = q_i \bigg(1-\prod_{j \in \mathcal{K}_i} (1-y_j(t-1))\bigg).
\label{eq:update}
\end{equation}
Equation \eqref{eq:update} follows from noting that the probability that $\sigma_j$ and $\tilde{\sigma_j}$ are equal is $(1-y_j)$ and thus the probability that all inputs to node $i$ are equal is the above product.  Note that this equation uses topological information contained in the $\mathcal{K}_i$.  However, we have treated ${y_j}$, $y_{j'}$ and $q_i$ as if they were probabilities of statistically independent random events.  We hypothesize that this semi-annealed protocol might be expected to yield good results for frozen-in cases when the network is large and the fraction of network nodes on short loops is small (the network is `locally tree-like').  To see the problem posed by short loops, consider a node with two inputs that themselves have inputs both coming from a common node; in this case, the elements of $y(t)$ in Eq. \eqref{eq:update} are no longer statistically independent and multiplying the probabilities is no longer correct.  See Ref. \cite{Ott2008} for discussion related to the locally tree-like assumption.  Our numerical tests of frozen networks indeed yield results that agree very well with our semi-annealed hypothesis on large, locally tree-like networks.  We also find our predictions to hold for networks with a large number of feedforward motifs, a nontree-like three-node subgraph that has been found to be prevalent in real gene networks \cite{Alon2002}.

The case where both network states are exactly the same corresponds to $y_i(t) = 0$, which is a fixed point of Eq. \eqref{eq:update}.  In order to determine the stability of this fixed point, we linearize Eq. \eqref{eq:update} around $y(t) = 0$ for small perturbations:
\begin{equation}
y_i(t+1) \approx q_i \sum_{j=1}^N A_{ij}y_j,
\label{eq:linupdate}
\end{equation}
where $A_{ij}$ are the elements of the adjacency matrix $A$.  Equation \eqref{eq:linupdate} can be written in matrix form as $y(t+1) = Qy(t)$ where 
\begin{equation}
\label{eq:Qmatrix}
Q_{ij} = q_i A_{ij}.
\end{equation}
The stability is thus governed by the largest eigenvalue $\lambda_Q$ of this matrix: 
\begin{eqnarray}
\nonumber \lambda_Q &>& 1, \text{$y = 0$ is unstable;} \\
\label{eq:stability} \lambda_Q &=& 1, \text{$y = 0$ is critical;} \\
\nonumber \lambda_Q &<& 1, \text{$y = 0$ is stable.}
\end{eqnarray}
Since $Q_{ij} \geq 0$, the Perron-Frobenius theorem \cite{MacCluer2000} guarantees that $\lambda_Q$ is real and positive.  We also note that, for any given adjacency matrix $A$ and assignment of $q_i$'s to nodes, Eq. \eqref{eq:update} can be iterated numerically to predict the expected time-asymptotic saturation value of the difference in two initially nearby states when evolved to steady-state.  We numerically test this prediction, as well as the stability criterion in Eq. \eqref{eq:stability} in the next section. \footnote{We note that Eq. \eqref{eq:Qmatrix} and the condition $\lambda_Q = 1$ also occurs in the treatment \cite{Ott2008} of site percolation on directed networks where different sites have different removal properties.  A similar condition involving $\langle K^2 \rangle / \langle K \rangle$ also arises in percolation on undirected networks \cite{Cohen}.}

As a special case of interest, if the the $q_i$ are uniform, $q_i \equiv q$, then $\lambda_Q = q\lambda$, where $\lambda$ is the maximum eigenvalue of the adjacency matrix.  This yields the critical condition,
\begin{equation}
\lambda = 1/q.
\label{eq:l_phase}
\end{equation}
Furthermore, for the case of a large network whose links are randomly assigned subject to a joint probability distribution $\tilde{P}(K^{in}, K^{out})$ at each node (with no assortativity), the mean field approximation for the largest eigenvalue is \cite{Ott2007}
\begin{equation}
\lambda \approx \frac{\langle K^{in}K^{out} \rangle}{\langle K\rangle},
\label{eq:meanfieldlambda}
\end{equation}
where, since $\langle K^{in}\rangle = \langle K^{out}\rangle$ necessarily, we use the notation $\langle K \rangle \equiv \langle K^{in}\rangle = \langle K^{out}\rangle$.  Equations \eqref{eq:l_phase} and \eqref{eq:meanfieldlambda} yield the same criterion as in Eq. \eqref{eq:corr_phase}.  In the case where $K^{in}$ and $K^{out}$ are uncorrelated, $\tilde{P}(K^{in}, K^{out}) = P_{in}(K^{in})P_{out}(K^{out})$ and $\langle K^{in}K^{out} \rangle = \langle K \rangle ^2$, yielding Eq. \eqref{eq:rbn_phase}.

The eigenvalue of random network adjacency matrices with assortativity has been considered in Ref. \cite{Ott2007}, which defines an assortativity measure $\rho$ as 
\begin{equation}
\label{eq:assortativity}
\rho = \frac{\langle K_i^{in}K_j^{out}\rangle_e}{\langle K^{in}K^{out}\rangle},
\end{equation}
where $\langle K_i^{in}K_j^{out}\rangle_e$ denotes an average over all links $(i, j)$ from node $i$ to node $j$.  The network is assortative (disassortative) if $\rho > 1$ ($\rho < 1$).  For $\rho$ near one, the largest eigenvalue $\lambda$ is approximately given by \cite{Ott2007}
\begin{equation}
\label{eq:asseigen}
\lambda \approx \frac{\langle K^{in}K^{out}\rangle}{\langle K \rangle}\rho.
\end{equation}
Thus by Eqs. \eqref{eq:update} and \eqref{eq:stability}, it is predicted that, for uniform $q$, assortativity (disassortativity) decreases (increases) the critical $q$ value.

In the case of nonuniform $q_i$, we have recently generalized Eq. \eqref{eq:meanfieldlambda} to obtain an analogous mean field approximation to $\lambda_Q$ without assortativity or community structure,
\begin{equation}
\label{eq:nonuniformq}
\lambda_Q \approx \frac{\langle q K^{in}K^{out}\rangle}{\langle K \rangle}.
\end{equation}
Our derivation of \eqref{eq:nonuniformq} will be published elsewhere.  From \eqref{eq:nonuniformq}, we see that correlation (anticorrelation) between $q$ and $K^{in}K^{out}$ decreases (increases) network stability and that, in the absence of correlation, the result is similar to that for a uniform $q$, $\lambda_Q \approx \langle q \rangle\langle K^{in}K^{out}\rangle/\langle K \rangle$, with $\langle q \rangle$ replacing the uniform $q$ (Eqs. \eqref{eq:stability} and \eqref{eq:l_phase}).

We now consider the generalization to allow any number of discrete node states.  We denote the number of possible states of node $i$ by $S_i$, and we label the possible states $0, 1, 2, ..., S_i-1$.  The number of possible inputs to $i$ from the set $\mathcal{K}_i$ of nodes that influence it is $\prod_{j \in \mathcal{K}_i} S_j$.  For each of these possible inputs, the truth-table function $f_i$ in Eq. \eqref{eq:definition} assigns one of the $S_i$ possible states to node $i$.  Similar to the Boolean case, we take the assignment to be random and to have an `expression bias' $p_{i,s}$ for each of the $s = 0, 1, 2, ... S_i-1$ node states, where $p_{i,s}$ denotes the probability that $f_i$, for a given set of inputs, assigns the state $s$ to node $i$, and $\sum_s p_{i,s} = 1$ for all nodes $i$.  As in the Boolean case, we can then introduce the sensitivity $q_i$ giving the probability that two different sets of inputs result in a different updated state of node $i$, which, in the random truth table case,
\begin{equation}
\label{eq:gensensitivity}
q_i = 1 - \sum_s p_{i,s}^2.
\end{equation}
With this definition, we see that all our previous reasoning still applies, and Eqs. \eqref{eq:update}-\eqref{eq:stability} hold with this generalized expression for the node sensitivities and with $y_i(t)$ interpreted as the probability of disagreement between $\sigma_i(t)$ and $\tilde{\sigma_i}(t)$.  In the case of uniform number of node states $S_i \equiv S$ and equal expression biases $p_{i,s} \equiv p_s = 1/S$, among these states, Eq. \eqref{eq:gensensitivity} becomes $q = 1 - 1/S$, which, when combined with Eq. \eqref{eq:l_phase} yields the previous result in Eq. \eqref{eq:multiphase}.

Finally, we note that our criticality criterion, $\lambda_Q = 1$, is unchanged by the presence of delays, as in the models of Refs. \cite{Ghil}, and only a slight modification is required of Eq. \eqref{eq:update} (i.e., $y_j(t-\tau_{ij})$ replaces $y_j(t-1)$).  The condition $\lambda_Q = 1$ implies that the components of $y$ in Eq. \eqref{eq:update} are time-independent.  Thus we predict that the delays $\tau_{ij}$ do not influence the result, and the criticality condition in Eq. \eqref{eq:stability} is independent of the synchronous update structure of the most commonly used random Boolean network models.  Similarly, the time-asymptotic steady state obtained by repeated iteration of \eqref{eq:update} is, by definition, time-independent and thus also does not depend on the $\tau_{ij}$ (although the $\tau_{ij}$ will influence the time-dependent approach to the asymptotic steady state; see supplementary material).

\section{\label{sec:numerical}Statistical Methods}

We numerically test the above predictions on several classes of Boolean networks with uniform sensitivity (i.e., $q_i = q$ is the same for all nodes):
\begin{enumerate}
\renewcommand{\labelenumi}{(\alph{enumi})}
\item random networks with $K^{in} = K^{out}$; 
\item random networks with imperfect correlation between $K^{in}$ and $K^{out}$; 
\item networks with assortativity or disassortativity; and
\item networks constructed as in (a) but with a substantial number of feedforward loops.
\end{enumerate}

We additionally test our predictions on two classes of networks with nonuniform sensitivities: 
\begin{enumerate}
\renewcommand{\labelenumi}{(\alph{enumi})}
\item[(e)] networks constructed as in (a) but where nodes have different sensitivities correlated with the degrees of the nodes; and 
\item[(f)] networks with significant community structure, where the two communities have different, uniform sensitivities.
\end{enumerate}
Finally, we test our generalization to more than two node states on networks of type (a) but with $S_i = 4$ for all nodes.  For types (a)-(c) and (e), we use networks with truncated power-law degree distributions.  (Evidence for the presence of this type of distribution in gene networks has been seen in \cite{Scalefree}.)

The algorithms for constructing the networks of types (a)-(c) are as follows.  (i) Establish the in- and out-degrees for each node, which are drawn from a distribution,
\begin{equation}
P(K) \propto  \begin{cases}
K^{-\gamma}, 						& K \leq K^{max},\\
0, 	& K > K^{max},
\end{cases}
\end{equation}
 where $\gamma = 2.1$ and $K^{max} = 15$ (Boolean case) or $K^{max} = 8$ ($S_i = 4$ case).  The out-degree is initially set to the in-degree.  (ii) Randomly swap the out-degrees between pairs of nodes.  If maximal correlation between in- and out-degrees is desired, as in (a), this step is skipped so that $K^{in} = K^{out}$ and $\langle K^{in}K^{out} \rangle$ is maximal.  A completely uncorrelated network has every nodal out-degree swapped exactly once, yielding $\langle K^{in}K^{out} \rangle = \langle K \rangle^2$.  The quantity $\langle K^{in}K^{out} \rangle$, which approximately determines $\lambda$ by Eq. \eqref{eq:meanfieldlambda}, can thus be tuned by the number of nodes that have their out-degrees swapped.  (iii)  Place links randomly between nodes subject to the constraints of the specified in- and out-degrees assigned at each node by the `configuration model' \cite{configmethod}.  (iv)  If networks with assortativity (disassortativity) are desired, as in (c), perform a given number of link swaps, as in \cite{Ott2007}, that increase (decrease) the assortativity $\rho$ in Eq. \eqref{eq:assortativity}.  In all cases we employ networks with $N = 10^4$ and two initial conditions separated by a Hamming distance of 100.  In the supplementary material we discuss finite size effects that can occur for smaller $N$.

We emphasize that, although we determine our networks randomly, in our numerical experiments we do not average over this randomness.  Rather, we generate one random network for each experiment and examine the resulting behavior of that specific network.

\section{\label{subsec:teststab}Results}

We test the steady-state predictions of Eq. \eqref{eq:update} and the criticality condition of Eq. \eqref{eq:stability} in Fig. \ref{fig:HvsQ}, and compare the calculated critical parameters to the mean-field-type approximations of Eqs. \eqref{eq:meanfieldlambda}, \eqref{eq:assortativity}, and \eqref{eq:nonuniformq} in the supplementary notes.  In order to compare Eq. \eqref{eq:update} (solid curves in Fig. \ref{fig:HvsQ}) to experimental measurements of the Hamming distance from numerical evolution of true frozen Boolean dynamical systems (markers in Fig. \ref{fig:HvsQ}), we calculate the node averaged steady-state fractional Hamming distance,
\begin{equation}
\label{eq:theory}
\bar{y} = \lim_{t \rightarrow \infty}\frac{1}{N} \sum_i y_i(t).
\end{equation}
In practice, this limit is calculated as the average Hamming distance from $t = 90$ to $t = 100$ when all delays are the same ($\tau_{ij} = 1$), and from $t = 490$ to $t = 500$ when nonuniform delays are present. These times are well after the steady-state value is reached (see supplementary material).  Each experimental data point in Fig. \ref{fig:HvsQ} corresponds to a single realization of interconnections averaged over 100 realizations of the time-independent truth table with specified sensitivity as before.  

\subsection{In-/Out-degree Correlations and Heterogeneous Time Delay}
Figure \ref{fig:HvsQ}(a) shows the steady-state Hamming distance as a function of the sensitivity for one network of type (a) ($\lambda = 4.4$) and two of type (b) ($\lambda = 2.9, 2.3$).  The closed markers in the figure represent experiments with uniform delay $\tau_{ij} = 1$ on all links, while the open markers correspond to experiments where half the links, randomly chosen, have $\tau_{ij} = 10$ and the remainder have $\tau_{ij} = 1$.  Importantly, the degree distributions are the same for all three networks, and we attain different $\lambda$ values by varying the correlation between the in-degree and the out-degrees.  We see from Fig. \ref{fig:HvsQ}(a) that there is close agreement between the theoretical prediction and the experimental results and our prediction that the presence of delays does not change the stability is confirmed.  Additionally, the measured steady-state Hamming distance is essentially zero below the critical value of the sensitivity, $q_{crit} = 1/\lambda$ (this point is indicated by vertical downward arrows in Fig. \ref{fig:HvsQ}(a)).  We emphasize that the degree distributions (and hence $\langle K \rangle$) are the same for the networks in Fig. \ref{fig:HvsQ}(a), and thus, if the in-/out-degree correlation were ignored, the observed difference between the stability conditions for these networks would not be predicted.

\subsection{Assortativity/Disassortativity}
Figure \ref{fig:HvsQ}(b) shows results obtained when significant assortativity or disassortativity is present (type (c) networks).  In this experiment, as well as all those reported below, the delays are all uniform.  The networks under consideration have the same joint degree-distribution with $K^{in} = K^{out}$.  However, each of the networks have very different assortativities ($\rho = 0.52, 1.0, 1.7$, defined in Eq. \eqref{eq:assortativity}), which yield different largest eigenvalues ($\lambda = 3.0, 4.4, 9.9$).  Since the joint degree distributions are the same, Eq. \eqref{eq:corr_phase} would predict that the three networks have the same stability characteristics.  However, since their eigenvalues are very different, we predict that, as observed, the transitions of the three networks occur at different values of $q$.  Again the theoretical predictions of $q_{crit}$ are indicated by vertical arrows.

\subsection{Motifs}
Random construction of networks, as used in the networks above, is expected to yield networks that are locally tree-like \cite{Ott2007}.  However, we note that biological and other types of networks often have motifs (small subgraphs) that occur with higher frequency than in randomly constructed networks \cite{Alon2002}.  For gene networks of \textit{E. coli} and \textit{S. cerevisiae}, it was found that the number of feedforward loop motifs (see inset to Fig. \ref{fig:HvsQ}(c)) is significantly enhanced compared to the expected number in a randomly constructed network.  In these networks, the number of feedforward loops per node $c$ is roughly 0.1.  Thus we consider a network of type (a) ($\lambda \approx 2.9$ and $N = 10^4$) after adding 1000 ($c = 0.1$) and, in an extreme case, 2000 ($c = 0.2$) feedfoward loops.  To add a feedforward loop, we randomly choose a node $A$, follow a random output to node $B$, and follow a random output of $B$ to node $C$.  We then add a link from node $A$ to node $C$.  We do this a given number of times, avoiding nodes that already participate in an added feedforward loop.  In Fig. \ref{fig:HvsQ}(c), we see that the semi-annealed theory of Eq. \eqref{eq:update} (solid curve) again agrees well with our numerical experiments (solid markers).  Based on such results, we believe that the locally tree-like network requirement does not invalidate application of our method to real gene networks.  We also note that the critical point is essentially unchanged by the addition of loops (adding links only slightly increases the largest eigenvalue), however more feedforward loops tend to increase the steady-state Hamming distance  for $q > q_{crit}$.

\subsection{Application to S. cerevisiae}
As a real biological example, we include in the supplementary notes a graph similar to those in Figs. \ref{fig:HvsQ}(a)-(c) using a published network for the yeast \textit{S. cerevisiae} \cite{Yeast}.

\subsection{Heterogeneous Correlated Sensitivities}
Figure \ref{fig:HvsQ}(d) demonstrates the effect of a distribution of $q_i$'s on the stability of a network with $K^{in}_i = K^{out}_i = K_i$ and with correlation between the nodal values of $q_i$ and $K_i$, i.e., type (d) networks.  We consider two situations, one where $\langle q K^2 \rangle/(\langle q \rangle\langle K^2 \rangle)$ is maximal, and one where it is minimal.  The $q_i$ are drawn from a uniform distribution centered at $q_0$ (the abscissa in the figure), with width $\Delta q = 0.1$.  Maximal (minimal) $\langle q K^2 \rangle$ is attained by assigning the largest $q_i$ to the node with the largest (smallest) $K_i$, the second largest $q_i$ to the node with the second largest (second smallest) $K_i$, and so on.  As can be seen from the figure, there is good agreement between the semi-annealed theory and the numerical experiments, and the two networks become unstable at different values of $q_0$.  (Vertical arrows again indicate the points where $\lambda_Q = 1$.)

\subsection{Community Structure}
Figure \ref{fig:HvsQ}(e) shows our results for a case where there is community structure and community-dependent sensitivity.  To construct the networks in Fig. \ref{fig:HvsQ}(e), consider the case where there are two communities, and we assign a link from node $i$ in community $a$ to node $j$ in community $b$ with probability $\theta_{ab}$.  We impose the additional constraints that $\theta_{aa} = \theta_{bb} \equiv \theta_\cup$ and that $\theta_{ab} = \theta_{ba} \equiv \theta_\cap$, and the size of the two communities are the same, $N/2$.  We take $\langle K^{in} \rangle = \langle K^{out} \rangle = \langle K \rangle = (\theta_\cup+\theta_\cap)N$ to be the same for both communities, and we also assume that communities $a$ and $b$ have different sensitivities $q_a$ and $q_b$, respectively.	As $\theta_\cap$ is increased from zero to $\theta_\cap = \theta_\cup$, $\lambda_Q$ changes from the case of two completely separated communities to one of a single random network.  Communities $a$ and $b$ have equal sizes of 5000 nodes, community $a$ has $q_a = 0.5$, and community $b$ has $q_b = 0.1$.  In order to vary $\lambda_Q$, we vary $\theta_\cup$ and $\theta_\cap$, keeping their sum constant in order to maintain constant $\langle K \rangle$.  As with the curves in Fig. \ref{fig:HvsQ}(a)-(c), the transition to chaos is governed by $\lambda_Q$ ($\lambda_Q$ = 1 at the vertical arrow), and Eq. \eqref{eq:update} (solid curve) accurately predicts the numerically observed (solid circles) steady-state Hamming distance.

\subsection{Non-Boolean Models}
\begin{figure}
\includegraphics[scale=.52]{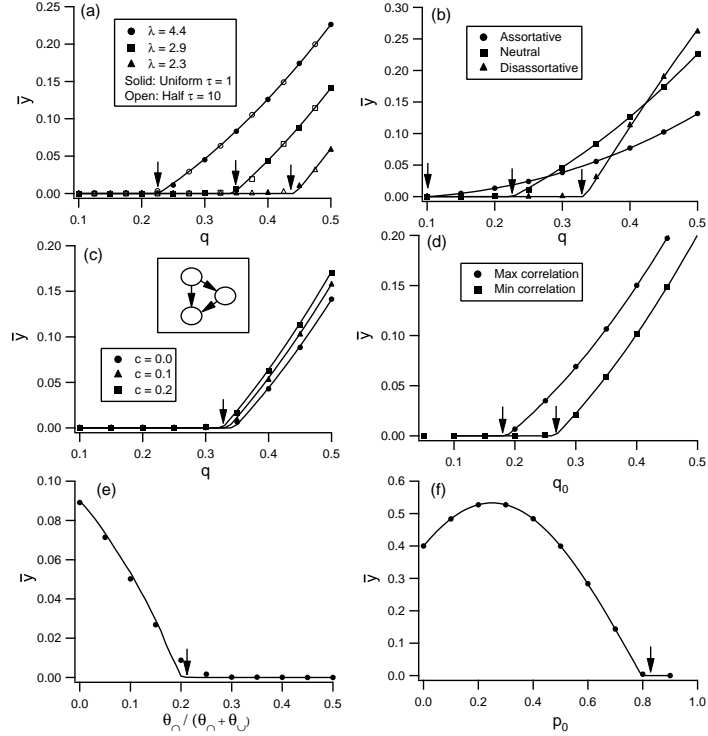}
\caption{\label{fig:HvsQ} (a) $\bar{y}$ vs. $q$ for three networks with different largest eigenvalues ($\lambda \approx 5.5, 3.4, 2.3$), both with uniform delay on all links $\tau_{ij} = 1$ (closed markers) and with half the links having increased delay of $\tau_{ij} = 10$ (open markers).  The solid curves correspond to the prediction $\bar{y}$ (defined in Eq. \eqref{eq:theory}) obtained by simulating Eq. \eqref{eq:update}.  The downward vertical arrows correspond to $q_{crit} = 1/\lambda$ for each of the three networks.  (b) $\bar{y}$ vs. $q$ for three networks with different assortativities.  (c) $\bar{y}$ vs. $q$ for networks with added feedforward motifs, with an illustration of the feedforward motif (inset).  (d) $\bar{y}$ vs. $q_0$ for networks with maximum correlation (circles) and minimum correlation (squares) between $K^{in}_i K^{out}_i$ and $q_i$, where $q_i$ is drawn from a uniform distribution centered at $q_0$ with width $0.1$. (e) $\bar{y}$ vs. $\theta_\cap / (\theta_\cup+\theta_\cap)$ for networks with community structure, where the two communities have $q_a = 0.5$ and $q_b = 0.1$. (f) $\bar{y}$ vs. $p_0$ for a network where each node can take one of $S_i = 4$ possible states.  $p_0$ is the probability that a zero appears in the truth table output; the remaining three symbols appear with equal probability.}
\end{figure}

Figure \ref{fig:HvsQ}(f) illustrates an application to a case in which there are more than two possible states at each node.  In particular, we consider $S \equiv S_i = 4$ possible states at each node.  (Since the number of possible inputs to the truth table for node $i$ in this case is  $4^{K^{in}_i}$, we take $K^{max} = 8$ due to memory constraints.)  Labeling the possible node states $\sigma_i \in \{0, 1, 2, 3\}$, we take nodes to have uniform expression biases for occurence of state-label 0, $p_0 \equiv p_{i,0}$, from 0 to 1. The three remaining labels ($\sigma = 1, 2, 3$) also have uniform biases for all nodes $i$, $p_s \equiv p_{i,s} = (1-p_{i,0})/3$.  From Eq. \eqref{eq:gensensitivity}, $q \equiv q_i = 1-[p_0^2+(1-p_0)^2/3]$, which has a maximum $q_{max} = 0.75$ at $p_0 = 0.25.$  As can be seen in the figure, the predicted fraction of nodes with differing states $\bar{y}$ (solid curve) also has a maximum there.  It is again seen that the measurements (markers) are well-predicted by the theory.


\section{\label{sec:discussion}Discussion}

In this paper, we have presented theoretical results (Eqs. \eqref{eq:update} and \eqref{eq:stability}) which predict the steady-state Hamming distance between states evolved from two nearby initial conditions and the stability of a given network.  These results are derived using the hypothesis that a theory derived in the semi-annealed case approximates the true situation, where by semi-annealed we mean that the network of connections is frozen, but the truth table at each node is randomly reassigned at each timestep.  For large networks, this approximation was found to give excellent agreement with the true case of frozen connections and frozen truth tables.  Our semi-annealed hypothesis does not rely on gross statistical properties of the network, but instead uses the specific network topology, as characterized by the network adjacency matrix, and the individual node sensitivities to make predictions.  

We tested our theoretical predictions with numerical experiments.  Previously unaddressed issues that we considered include the effects of assortativity, nonuniform time delay, nonuniform sensitivity, motifs, and community structure.  In all cases tested we found good agreement with our theory.

The theory that we have presented and tested above may represent a step forward in facilitating the application of discrete state dynamical network models to biological systems.  Given a specific genetic interaction network and an estimate of the node sensitivities, Eq. \eqref{eq:stability} predicts the stability of that particular network directly from the adjacency matrix.  Curated networks already exist in the literature for model single-cellular systems, and new algorithms continue to be developed for inferring interaction networks from a wide range of data sources (microarray experiments, GO annotation, genome sequencing, etc.). 
We note that such a procedure has the advantage that, because the actual experimentally determined network is employed, topological aspects such as nodal in-/out- degree correlation, assortativity, community structure, etc., do not first have to be determined and then statistically modeled.  Thus, by use of our stability criterion \eqref{eq:stability}, there is the potential that future analysis may be able to evaluate a supposed relationship between the stability characteristics of various networks and their functioning.  For example, one might test whether cancer gene networks are less stable than those in healthy tissue.  This could lead to the strong variations in gene expression observed in cancerous tissue \cite{w1}, even when the underlying gene network is unchanged.  We are currently pursuing research along this line.  

This work was supported by NSF (Physics) and by ONR (contract N00014-07-1-0734).  We thank L. Staudt for discussion.  The work of A.P. was partly supported by the NCI intramural program.

\section{Supplementary Information 1: Transient Evolution}

\renewcommand{\thefigure}{S\arabic{figure}}
\setcounter{figure}{0}

\begin{figure}
\includegraphics{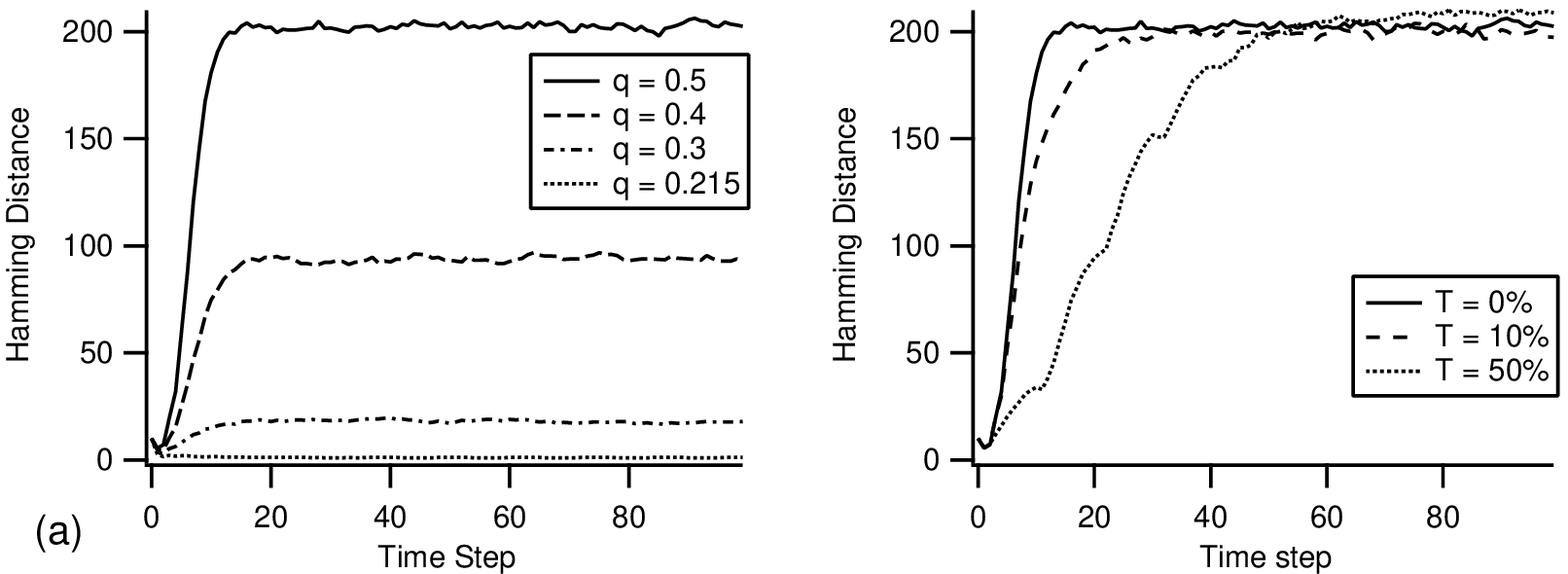}
\caption{\label{fig:HvsT} (a) Evolution of the Hamming distance between two initial conditions for a typical network of size $N=1000$ and $\epsilon = 0.01$ for various values of the sensitivity and uniform delay $\tau_{ij} = 1$.  (b) Evolution of the Hamming distance between two initial conditions for a typical network of size $N=1000$, $q = 0.5$, and $\epsilon = 0.01$.  Results are shown for a network with $\tau_{ij} = 1$ for all links (solid curve) and with $\tau_{ij} = 10$ on 0.1 (dashed curve) and 0.5 (dotted curve) of the links.}
\end{figure}

Figure \ref{fig:HvsT} shows the time evolution results for networks of type (a) (i.e., $K^{in}_i = K^{out}_i$ for all $i$ and uniform $q_i = q$).  Once a random network is generated, we simulate the evolution of two close initial conditions and plot the Hamming distance as a function of time in Fig. \ref{fig:HvsT}. Specifically, we take an arbitrary initial condition and generate a perturbed initial condition by flipping a fraction $\epsilon$ of the state bits; in Fig. \ref{fig:HvsT}, $\epsilon = 0.01$ and $N = 1000$ corresponding to 10 flipped bits.  Figure \ref{fig:HvsT}(a) shows the Hamming distance as a function of time step $t$ for four cases with different values of sensitivity ($q = 0.5, 0.4, 0.3, 0.2$) and uniform delays $\tau_{ij} = 1$ (as in Eq. [4]).  Each of these four curves are generated using the same network of interconnections (for which $\lambda = 4.3$) and the same perturbation in initial conditions averaged over 100 realizations of the nodal truth tables.  For the three cases $q = 0.5, 0.4, 0.3$, $\lambda_Q = q\lambda > 1$, the network is predicted to be unstable.  We see in Fig. \ref{fig:HvsT}(a) that in these cases, the Hamming distance rises and eventually saturates at a constant value.  In the fourth case, $q = 0.215$ and we have that $q\lambda < 1$, and the network is predicted to be stable, which is demonstrated in the figure. These cases illustrate the strong effect that in-/out-degree correlations can have: for the value $\langle K \rangle = 1.89$ in the network of Fig. \ref{fig:HvsT}, the prediction from the result for uncorrelated networks, Eq. [1], is stability for all values of $q$ (the minimum value of $1/q$, the right hand side of [1], is 2, which exceeds $\langle K \rangle = 1.89$).  Figure \ref{fig:HvsT}(b) shows time traces of the Hamming distance for $q = 0.5$ when non-uniform delays are present.  In the curves shown, a fraction $T = 0, 0.1, 0.5$ of the links are randomly chosen and given delays of $\tau_{ij} = 10$ with the remaining links having delay $\tau_{ij} = 1$.  The curves are for the same network as in Fig. \ref{fig:HvsT}(a) with $q = 0.5$ and are again the average of 100 different realization of the truth table.  (The choice of delayed links is the same for all 100 realizations.)  In each case, we see that the network is unstable and the Hamming distance rises to the same steady-state value, albeit at a slower rate for larger $T$.  This result thus is consistent with our prediction that whether or not a network is stable and its final saturation value do not depend on heterogeneity of the delays.

\section{Supplementary Information 2: Finite-Size Effects}

\begin{figure}
\includegraphics[scale=0.8]{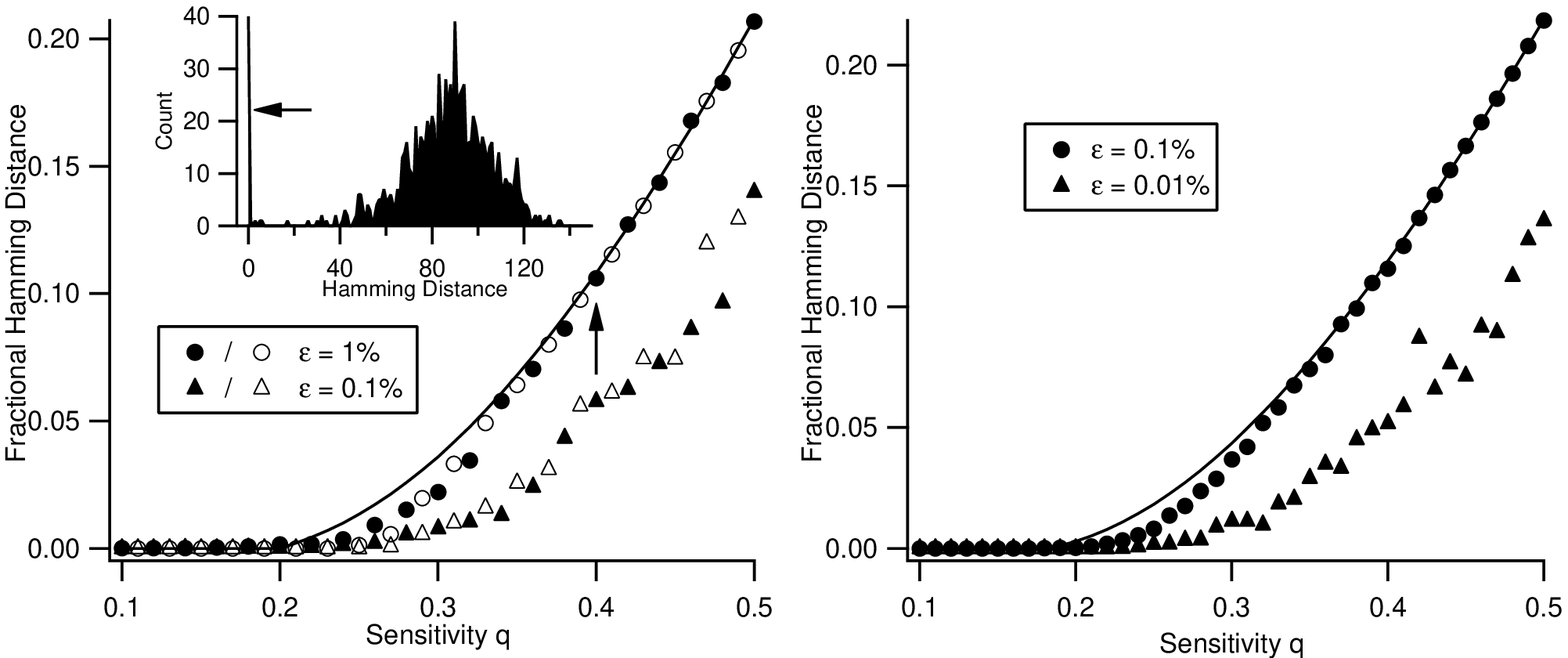}
\caption{\label{fig:finitesize} The steady-state fractional Hamming distance $h/N$ for (a) $N=10^3$ and (b) $N=10^4$ as a function of the sensitivity $q$ for various values of $\epsilon$, both in the frozen case (filled symbols) and the annealed case (open symbols).  The largest eigenvalue of this network's adjacency matrix is $\lambda \approx 5$.  While the theory does not depend on the value of $\epsilon$, finite-size effects cause a dependence on the number of flipped bits.  The inset to (a) shows a histogram of measured Hamming distances at $q = 0.4$ and $\epsilon = 0.01$ (up arrow).}
\end{figure}

In Fig. \ref{fig:finitesize}(a) and (b), we consider the importance of finite-size effects by varying $\epsilon$ (a parameter which does not appear in the theory) for two different size networks of type (a).  Figure \ref{fig:finitesize}(a) also compares the results of simulating the frozen case (solid markers) to the semi-annealed case described in the Theory section (open markers) for $N = 10^3$.  As before, in simulating a semi-annealed network, at each time step the nodal truth tables are randomly generated with the same $q$.  The networks under consideration in Figs. \ref{fig:finitesize}(a)-(b) have $T = 0$, $\lambda \approx 5.0$, and (a) $N = 10^3$ and (b) $N=10^4$.  As the figure demonstrates, larger $\epsilon$ yields better agreement with the theory, and the semi-annealed case seems indistinguishable from the frozen case.  Note also that the results for $\epsilon = 0.01$ ($\epsilon = 10^{-3}$) and $N = 10^3$ is similar to that for $\epsilon = 10^{-3}$ ($\epsilon = 10^{-4}$) and $N = 10^4$, suggesting that the relevant quantity is $\epsilon N$, the number of flipped states.  The inset of Fig. \ref{fig:finitesize}(a) shows the histogram of the Hamming distances used in calculating the point $q = 0.4$, $\epsilon = 0.01$ (upward vertical arrow).  The different trials used in generating this histogram correspond to different truth table realizations.  The distribution shown in the inset consists of a large number of samples with Hamming distance zero and a roughly symmetric part that has a mean near the theoretical prediction.  The overestimation of the mean by the theory therefore seems to be driven by the relative number of zero samples compared to the symmetric part.  

In order to understand the origin of the zero samples, we note that one way that they can arise is through `irrelevant' nodes (i.e., nodes that do not influence the dynamics of the network) and `frozen' nodes (i.e., nodes whose output is independent of its inputs due to the random assignment of the truth table).  Irrelevant nodes can arise by either having no out-going links or by inputting only to other irrelevant or frozen nodes.  Flipping the value of an irrelevant node, by definition, does not change the subsequent evolution of the network; if a perturbation between nearby initial conditions consists solely of such flips, that perturbation dies out quickly.  Assuming that the fraction of irrelevant nodes is independent of $N$, then the probability that all $\epsilon N$ nodes for which the two initial conditions differ are irrelevant goes to zero as $N \rightarrow \infty$ for constant $\epsilon$; in this case, the observations should exactly match the theoretical prediction.  This is consistent with the trend indicated by our comparison of the $N = 10^3$ network in Fig. \ref{fig:finitesize}(a) with the $N = 10^4$ network in Fig. \ref{fig:finitesize}(b).

\section{Supplementary Information 3: Comparison of Mean-field Eigenvalue Approximations with Exact $\lambda_Q$}

For the systems tested in our numerical experiments, Table \ref{tab:compare} shows the critical parameter values at the stability/instability border as obtained from direct calculation of the maximum eigenvalue of the matrix $Q$ for the relevant specific networks (downward arrows in Fig. 1 (a)-(f)) compared to the corresponding results predicted from the mean-field-type theoretical estimates (Eqs. [11], [13]-[14], and [18]). For the community structure example we use the approximation,
\begin{equation}
\lambda_Q \approx \frac{N}{2}\bigg\{\theta_\cup(q_a + q_b) + \big[ \theta_\cup^2(q_a - q_b)^2 + 4\theta_\cap^2q_aq_b\big]\bigg\},
\label{eq:communities}
\end{equation}
which applies for the case of two equal communities with symmetric connectivity probabilities ($\theta_{ab} = \theta_{ba} = \theta_\cap, \theta_{aa} = \theta_{bb} = \theta_\cup$) as in Fig. 1(e).  The analysis leading to \eqref{eq:communities} will be published elsewhere.

The largest eigenvalue approximations predict the observed transition to unstable behavior quite well, as seen in the table below.  The only exceptions to this agreement are in the case of significant assortativity or disassortativity; however, this is to be expected since the approximate theory is a linear approximation for values of $\rho$ close to one.  The values of assortativity and disassortativity used in the paper (1.7 and 0.5) are far from this regime.  Nevertheless, even for these cases, the theory correctly predicts the qualitative trend that assortativity (disassortativity) decreases (increases) the critical $q$.

 \renewcommand{\thetable}{S\arabic{table}}

\begin{table}[ht]
\caption{Comparison between criticality conditions evaluated directly from $Q$ and from the approximate theory.}
\begin{tabular}{@{\vrule height 10.5pt depth4pt  width0pt} c | c c }
\hline
\hline
& Direct Evaluation & Approximate Theory\\
Critical $q$'s from Fig. 1(a) & 0.22 & 0.23 \\
(Approx. Theory from Eq. [11]) &  0.34 & 0.34 \\
 & 0.43 & 0.44 \\
\hline
Critical $q$'s for Fig. 1(b)  & 0.10 & 0.13 \\
(Approx. Theory from Eq. [12]) & 0.22 & 0.23 \\
 & 0.33 & 0.45 \\
\hline
Critical $q$'s for Fig. 1(c) & 0.33 & 0.33 \\
(Approx. Theory from Eq. [11]) & 0.34 & 0.33  \\
 &  0.34 & 0.34 \\
\hline
Critical $q_0$ for Fig. 1(d) & 0.19 & 0.20 \\
(Approx. Theory from Eq. [14]) & 0.27 & 0.28 \\
\hline
Critical $\theta_\cap / (\theta_\cup+\theta_\cap)$ for Fig. 1(e) &  0.21 & 0.21 \\  
(Approx. Theory from Eq. [18]) \\
\hline
Critical $p_0$ for Fig. 1(f) & 0.80 & 0.80 \\
(Approx. Theory from Eq. [15]) \\
\hline
\end{tabular}
\label{tab:compare}
\end{table}

\section{Supplementary Information 4: Application to the Regulatory Network of S. cerevisiae}

\begin{figure}
\includegraphics{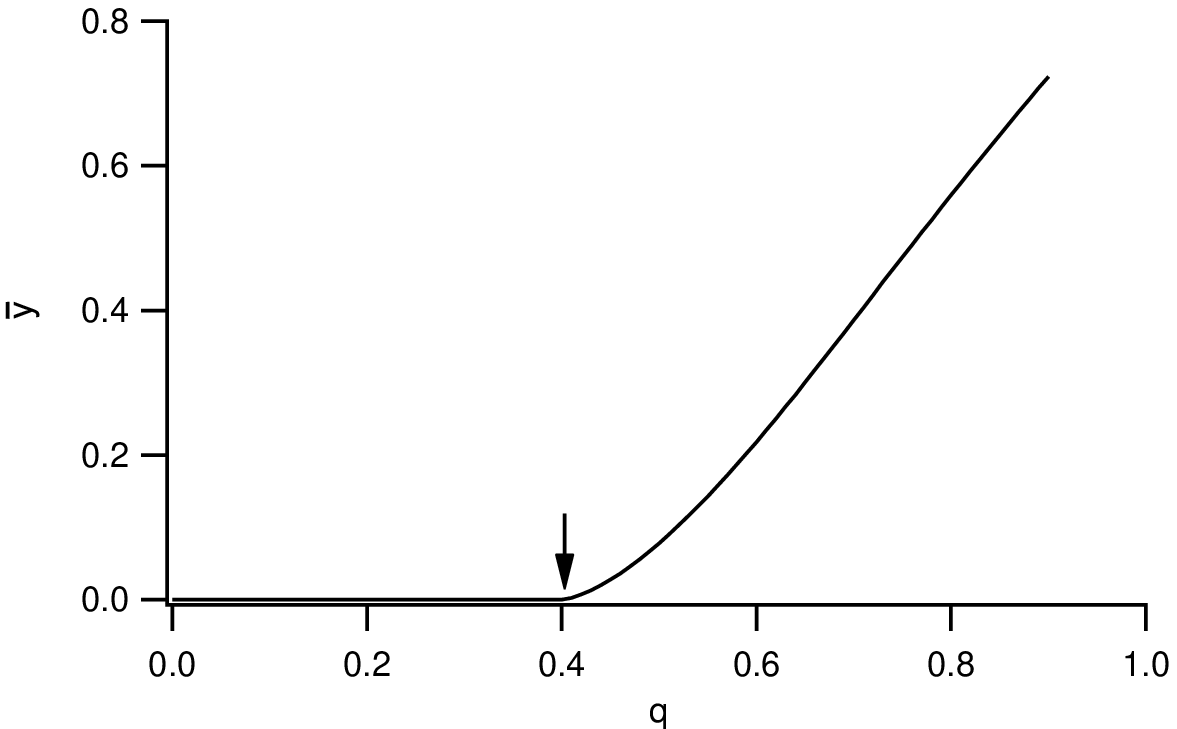}
\caption{\label{fig:yeasty} $\bar{y}$ vs. $q$, calculated from Eq. [17], for the published regulatory network of \textit{S. cerevisiae} [35].  The network undergoes a transition from stable to unstable behavior at $q_{crit} = 1/\lambda = 0.40$.}
\end{figure}

Figure \ref{fig:yeasty}, similar to Figs. 1(a)-(c), illustrates an application of Eq. [17] to the published network of the yeast \textit{S. cerevisiae} [35].  The largest eigenvalue of this network is $\lambda = 2.5$.  We have assumed in this plot that each node has the same sensitivity $q$, and again we see that the network undergoes a transition from stable to unstable behavior at $q_{crit} = 1/\lambda$.  However, in order to draw any conclusions about the criticality of the yeast regulatory network, we need a reliable estimate of the individual $q_i$'s, from which we can calculate $\lambda_Q$.  While estimating the sensitivities may be possible with existing microarray datasets, this is beyond the scope of this paper.

\end{document}